\documentclass[referee]{aa}
\usepackage{graphicx}
\usepackage{txfonts}
\usepackage{natbib}

\begin{document}

\title{Hall cascades versus instabilities in neutron star magnetic fields}
\titlerunning{Hall cascades versus instabilities in neutron star magnetic fields}

   \author{C. J. Wareing
           \and
           R. Hollerbach
           }

   \institute{Department of Applied Mathematics, University of Leeds,
              Leeds, LS2 9JT, UK.\\
              \email{cjw@maths.leeds.ac.uk}}

   \date{Received 12 October 2009 /
          Accepted 2 December 2009}

   \authorrunning{Wareing \& Hollerbach}

  \abstract
   {The Hall effect is an important nonlinear mechanism affecting the evolution
   of magnetic fields in neutron stars. Studies of the governing equation,
   both theoretical and numerical, have shown that the Hall effect proceeds in a 
   turbulent cascade of energy from large to small scales.}  
   {We investigate the small-scale Hall instability conjectured to exist from the
   linear stability analysis of Rheinhardt and Geppert.}
   {Identical linear stability analyses are performed to find a suitable
   background field to model Rheinhardt and Geppert's ideas. The
   nonlinear evolution of this field is then modelled using a three-dimensional
   pseudospectral numerical MHD code. Combined with the background field, 
   energy was injected at the ten specific eigenmodes with the greatest positive 
   eigenvalues as inferred by the linear stability analysis.}
   {Energy is transferred to different scales in the system, but not 
   into small scales to any extent that could be interpreted as a Hall 
   instability. Any instabilities are overwhelmed by a late-onset turbulent
   Hall cascade, initially avoided by the choice of background field, but
   soon generated by nonlinear interactions between the growing eigenmodes.
   The Hall cascade is shown here, and by several authors elsewhere, to
   be the dominant mechanism in this system.}
   {}


   \maketitle
%

\section{Introduction}

The Hall effect is now acknowledged to be an important mechanism in
the evolution of magnetic fields in the crusts of neutron stars.  As
derived by \cite{goldreich92}, the equation governing the magnetic field
under the influence of both the Hall effect and Ohmic diffusion is
\begin{equation}
\frac{\partial {\bf B}}{\partial t} = \nabla^2 {\bf B}
 - \nabla \times \left[ (\nabla \times {\bf B}) \times {\bf B} \right]
\label{eq:A}
\end{equation}
in which length is scaled by some characteristic length such as the depth $d$
of the crust, time scaled by the Ohmic decay time $4 \pi \sigma
d^2 / c^2$, where $\sigma$ is the conductivity, and $c$ is the speed of light.
The magnetic field is defined so that the Ohmic term,
$\nabla^2 {\bf B}$, and the Hall term,
$-\nabla\times\left[(\nabla\times{\bf B})\times{\bf B}\right]$, are
formally of the same order; this is accomplished by scaling $\bf B$ by
$nec/\sigma$, where $n$ is the electron number density, and $e$ is the
electron charge.  We note though that the true magnetic field in neutron
stars is typically several orders of magnitude greater than this, and
correspondingly the Hall timescale is several orders of magnitude faster
than the Ohmic timescale.  If $B_0$ is the non-dimensional amplitude of
$\bf B$ in Eq. (\ref{eq:A}), and $t$ is measured on the Ohmic timescale, then
$t'=B_0 t$ is measured on the Hall timescale.  We use both timescales
in this work, as appropriate.

Arguing by analogy with ordinary, hydrodynamic turbulence, Goldreich and
Reisenegger then conjectured that Eq. (\ref{eq:A}) would generate a turbulent Hall
cascade, transferring energy from large to small scales, with an energy
spectrum $E_k \propto k^{-2}$ (where $k$ is wavenumber), and a dissipative
cutoff occurring at $k \sim B_0$.  They also suggested that the total
energy could decay on the fast Hall timescale rather than the slow Ohmic
timescale -- despite the Hall term conserving energy and thus by itself 
being unable to cause decay on any timescale.
Instead, by transferring energy from large to small scales, the Hall
cascade enhances the efficiency of Ohmic decay, conjecturally by enough for
the total energy to decay on the fundamentally different, and faster
timescale.

Following Goldreich and Reisenegger's seminal work, numerous authors have
studied Eq. (\ref{eq:A}) theoretically and numerically, in both the original
spherical-shell geometry \citep{hollerbach02,hollerbach04,cumming04,pons07},
and the cartesian box geometries
\citep{biskamp96,biskamp99,dastgeer00,dastgeer03,cho04,shaikh05,cho09}
usually used in turbulence studies (because they allow much higher
resolutions than more complicated geometries).  The studies in box
geometries in particular all found spectra that are similar to the classical
5/3 Kolmogorov spectrum, as well as changes in slope that were interpreted
as a dissipative cutoff.  Results in two and three dimensions were also
found to be broadly similar. The turbulent Hall cascade has been
found to reach a stable equilibrium on a timescale of $t'\sim0.3-0.5$ 
\citep{cho04,cho09}, efficiently transferring energy from large to small
scales.

Our recent work \citep{wareing09a,wareing09b} has highlighted a
limitation of previous box simulations, in that they all employed
hyperdiffusivity, replacing the Ohmic term by $(\nabla^2)^\eta{\bf B}$,
where $\eta$ is typically 2 or 3.  As we noted, this masks the
equivalence of the terms in the governing equation. Both contain the
same number of derivatives so it is conceivable that the nonlinear
term will always dominate, even on arbitrarily short lengthscales. As 
demonstrated by \cite{hollerbach02}, one obtains a dissipative cutoff only
if one assumes that the cascade is local in Fourier space.

The argument is as follows: the ratio of the Hall
term to the Ohmic term is given by the field strength $B_0$,
independent of any lengthscales.  Implicitly a dependence on lengthscales 
may still exist: if the coupling is purely local in Fourier space, then the
relevant field strength is only the field {\it at that wavenumber}.
For sufficiently large $k$, this local field is then reduced sufficiently
for the Ohmic term to dominate the Hall term, resulting in a dissipative
cutoff at that $k$.  It is clear however how crucially this argument
depends on the coupling being purely local in Fourier space; if this is
not the case, then the same global $B_0$ applies to all lengthscales,
and the Hall term always dominates the Ohmic term.

Our 3D simulations \citep{wareing09b} reach a stable equilibrium by $t'\sim0.2$
and produce a smooth energy spectrum extending over the whole range of Fourier
space.  For large $B_0$, this tends towards the $E_k \propto k^{-2}$ scaling
suggested by Goldreich and Reisenegger.  We found no evidence, in either
2D or 3D, of a dissipative cutoff, implying that the Hall term is able
to dominate on all scales and that the coupling is nonlocal in
Fourier space. Additional evidence of the nonlocal nature of the Hall
cascade comes from the strong anisotropy in the presence of a uniform field
found by ourselves and others \citep{cho04,cho09}; if the coupling were purely
local in Fourier space, then including a uniform field would have no effect at
all on small scales, in contrast to what is observed.

In a very different approach, \cite{rheinhardt02} (henceforth referred
to as R\&G) performed a linear stability analysis of Eq. (\ref{eq:A}),
and showed that for a particular choice of background field, growing eigenmodes
exist at small wavenumbers $0 < k_x, k_y < 5$. They conjectured
that the transfer of magnetic energy from a background (large-scale)
field to small-scale modes may therefore proceed in a non-local way in
phase space, resulting in a Hall instability.  This instability could
be identified on the basis of its energy spectrum that does not decline
monotonically (e.g. as in a turbulence spectrum) but instead exhibits an
increasingly large peak at some large $k$, corresponding to a transfer of energy
directly from the largest scale to this small-scale peak.

We note that no calculation to date shows any 
such peak, so it already seems very likely that any these instabilities
are simply overwhelmed by the turbulent Hall cascade.  Nevertheless, we 
test this idea here in greater detail, by carefully selecting the initial
conditions to favour the development of a Hall instability, and inhibit
the development of a Hall cascade (at least initially).  However, we find
that even under these optimised circumstances, there is no evidence
that Hall instabilities play any significant role in Eq. (\ref{eq:A}).

\section{The R\&G linear stability analysis}

R\&G begin by considering a large-scale background field
${\bf B}_0$, which has no Hall term, that is, it must satisfy
\begin{equation}
\nabla\times[(\nabla\times{\bf B}_0)\times{\bf B}_0]={\bf0},
\label{ZeroHall}
\end{equation}
which is of course already a rather restrictive assumption.
They next linearise Eq. (\ref{eq:A}) about this background field to obtain
\begin{equation}
\frac{\partial {\bf b}}{\partial t} = \nabla^2 {\bf b}
 - \nabla \times \left[ (\nabla \times {\bf B}_0) \times {\bf b} +
(\nabla \times {\bf b}) \times {\bf B}_0 \right],
\label{eq:B}
\end{equation}
describing the behaviour of small perturbations {\bf b}.  If one
finally ignores the very gradual Ohmic decay of ${\bf B}_0$, and
instead treats it as being constant in time, then Eq. (\ref{eq:B}) becomes a
standard linear eigenvalue problem, with solutions that either
decay or grow exponentially.

Furthermore, whereas in Eq. (\ref{eq:A}) the Hall term conserves magnetic
energy $\int|{\bf B}|^2dV$, in Eq. (\ref{eq:B}) the now two linearised Hall
terms do not conserve $\int|{\bf b}|^2dV$.  It is indeed possible
therefore to obtain exponentially growing solutions, corresponding
to a transfer of energy from the background field to the perturbation.
What we wish to consider in this work is the subsequent nonlinear
evolution of these perturbations, including also the no longer
constant-in-time background field.

R\&G consider a plane layer geometry, periodic in $x$ and $y$ and
bounded in $z$, with either vacuum or perfectly conducting boundaries
at the top and bottom.  For their background field, they specify
${\bf B}_0 = f(z)\,{\bf e}_x$.  This satisfies Eq. (\ref{ZeroHall}) for any
choice of $f(z)$, and also has the additional advantage of decoupling
the horizontal wavenumbers $k_x$, $k_y$ for $\bf b$ (because ${\bf B}_0$
is independent of $x$ and $y$).  Equation (\ref{eq:B}) has therefore been
reduced to a linear, one-dimensional eigenvalue problem, in which only the
$z$ structure still has to be solved.

For suitable choices of $f(z)$ in particular with at least quadratic
curvature, so that the second derivative $f''\neq0$, they then find
that one can indeed obtain exponentially growing modes $\bf b$, that
is, instabilities of the large-scale field ${\bf B}_0$.  However, these
instabilities only occur for horizontal wavenumbers $k_x$ and $k_y$ up
to around 5 or so, which is nowhere nearly large enough to be considered
truly small-scale. The $z$ structure is also not really small-scale,
except for a narrow boundary layer that forms at large $B_0$. On the 
scale of the whole neutron star, the crust is only a thin layer, so the large
scales of R\&G could already be considered to be relatively small. But in the 
context of Hall cascades versus instabilities, which is of interest here,
R\&G's own linear stability analysis simply does not present any evidence
of a direct transfer from large to genuinely small scales.

Furthermore, even for these moderate-scale modes that do grow, the growth
rate is rather small, always less than 0.6 when measured on the Hall
timescale $t'$.  It would take several Hall timescales therefore for these
instabilities to grow by any appreciable amount.  In contrast, the
traditional Hall cascade is so efficient that $t'\sim0.2$ is already
enough to establish the full turbulence spectrum.

\section{Our linear stability analysis}

\begin{figure}
\centering
\includegraphics[width=8.8cm]{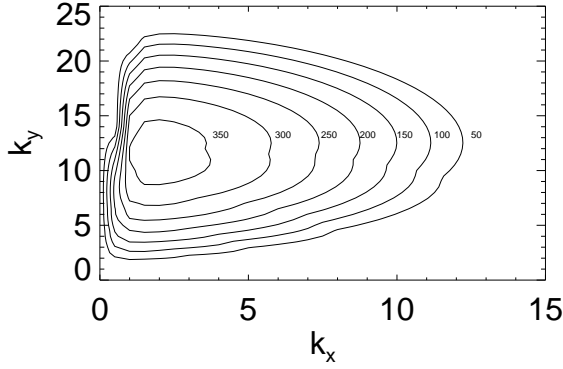}
\caption{Growth rates for those $k_x, k_y$ combinations that yield
exponentially growing modes, for $f(z)=1000[1+\sin(2z)]$.}
\label{eigenmodes}
\end{figure}

Our previously used 3D code \citep{wareing09b} is periodic not just in $x$
and $y$, but in $z$ as well, extending in all three directions
from $-\pi$ to $+\pi$.  Our function $f(z)$ must therefore also be
periodic, but otherwise exactly the same linear stability analysis as applied
by R\&G can be applied in this case.  
After experimenting with various choices, we found that
whilst functionally dissimilar to R\&G, the form 
$f(z)=B_0[1+\sin(2z)]$ yielded results that are qualitatively similar.
Forms of $f(z)$ that qualitatively very closely resemble the choices 
of R\&G, e.g., $f(z)=B_0[cos(z/2)]$ compared to R\&G's $f(z)=B_0[1-z^2]$, 
have not enabled us to identify unstable eigenmodes in our analysis.
Figure \ref{eigenmodes} shows the results for $B_0=1000$
(the same value as used by R\&G).  The highest growth-rate, corresponding to our
selection criteria for $f(z)$, is $\sim400$,
at $k_x=2$ and $k_y=11$, comparable to R\&G's peak value of $\sim550$, at
$k_x=1, k_y=0$ (with both growth-rates measured on the Ohmic timescale
$t$). For $k_y=0$, the most unstable eigenmodes are 
non-oscillatory; for $k_y>0$, they are oscillatory. This is the 
pattern also obtained by R\&G.

The main difference between our results and R\&G's is that ours extend
over a wider range of $k_x, k_y$ values.  This of course corresponds to
shorter lengthscales than they obtained, which if anything should then enhance
the Hall instability mechanism.  As we, however, now show there is 
no evidence that these `instabilities' play any significant
role in the dynamics of Eq. (\ref{eq:A}).

\section{Nonlinear evolution}

To study the nonlinear evolution of these eigenmodes, we considered the ten
most rapidly growing ones, and adjusted the energy in each to be
either 1\% or 10\% of the energy in the background field ${\bf B}_0$.
In total, the energy that is initially in the perturbations is therefore
either 10\% or 100\% of the energy in the background field.  These two
setups, background field plus either small or large perturbations, were
then used as the initial conditions in the original, nonlinear Eq. (\ref{eq:A}).
Both simulations were performed at resolutions of $128^3$ and $256^3$, with no
difference in the results.

\begin{figure*}
\centering
\includegraphics[width=18cm]{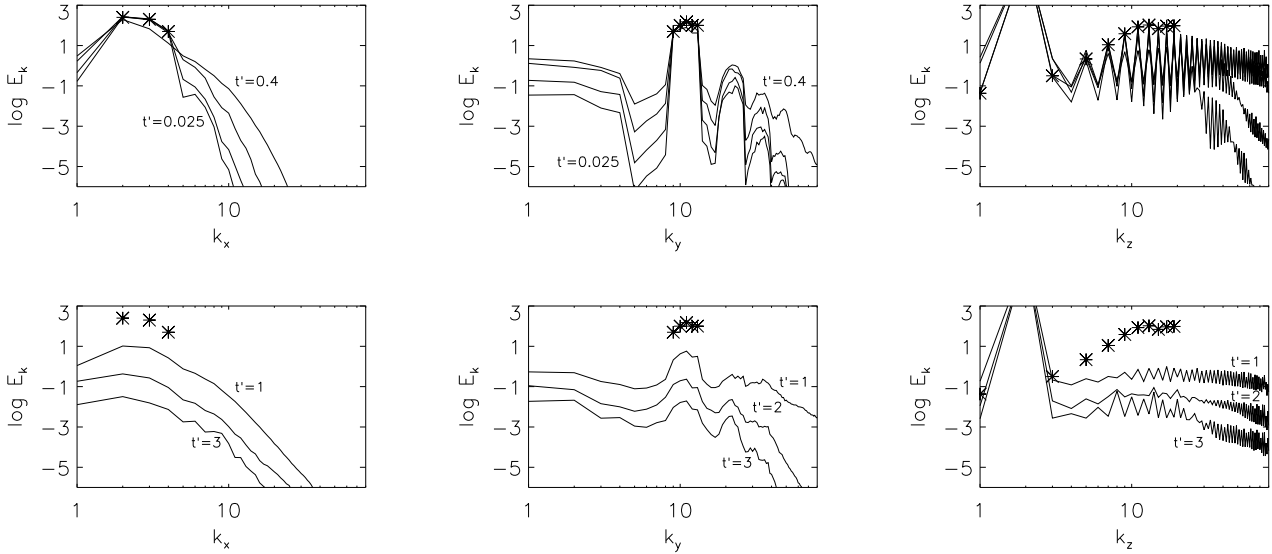}
\caption{The evolution of the field with a small perturbation. Power
spectra are shown at early times $t'=0.025, 0.1, 0.2$ and $0.4$ in 
the top row and at late times $t'=1, 2$ and $3$ in the bottom row. The
total energy is collapsed onto the $k_x$ axis (left), $k_y$ axis (middle),
and $k_z$ axis (right). Asterisks indicate the initial conditions at
$t=t'=0$.}
\label{small}
\end{figure*}

Figure \ref{small} shows the results for the small perturbations, 1\%
energy in each of the top ten eigenmodes.  At very early times, the
energy in these modes does indeed grow, and at the rates predicted by
the linear stability analysis.  However, this phase is so short, only up
to $t'\approx0.01$, that there is virtually no growth in this
time; with growth rates of $\sim0.4$ on this Hall timescale, the
perturbations grow by only a factor of $\exp(0.01\cdot0.4)=1.004$.
One could of course make this linear growth phase much longer, simply
by assuming the initial perturbations to be smaller. However, as soon
as they approach the $\sim1$\% energy level, the linear growth phase
ends, and one is once again in the regime shown here.

As indicated in Fig. \ref{small}, by $t'\approx0.025$, the
nonlinear interactions among these modes are clearly beginning to spread
the energy to different $k_x, k_y$ combinations.  This is most easily
seen in the $k_y$ spectrum, where the $k_y\approx10$ initial condition
generates higher harmonics at e.g., $k_y\approx20, 30$.  In the $k_x$
spectrum, the higher harmonics of the $k_x=2, 3, 4$ initial conditions
immediately blend together to form the beginning of the standard Hall
cascade.

There are two points to note regarding the $k_z$ spectrum.
First, the very strong peak (off the scale) at $k_z=2$ is the $\sin(2z)$
component of the background field.  Second, the perturbations have a
particular symmetry in $z$, containing only odd $k_z$.  Because the
background field contains only $\sin(2z)$, but no sine or cosine component of just $z$,
the perturbations decouple into even/odd $k_z$, and the odd $k_z$ modes
turn out to be the more unstable.  This initial condition of only odd
$k_z$ in the perturbations is the origin of the `zig-zag' pattern in the
$k_z$ spectrum.

As time progresses, the spectra then evolve exactly as one might expect
based on the standard Hall cascade picture.  For example, the initially
distinct peaks of the higher harmonics in $k_y$ are increasingly smoothed
out to form the standard turbulent cascade.  We note also the existence of an
inverse cascade, in which the regime $k_y<10$ is quite effectively filled in.

By $t'=3$, the memory of the particular initial condition has been largely
erased, and one is left simply with the standard cascade.  There is
certainly no evidence of any growing peaks, either at the moderate scales
of the original instabilities, or the small scales speculated by R\&G.
At even later times, the solutions eventually simply decay, again not
exhibiting any growing peaks at any particular lengthscale.  In
\cite{wareing09b}, solutions were obtained all the way to $t'=15$, still with
no peaks emerging from the turbulent cascade.

\begin{figure*}
\centering
\includegraphics[width=18cm]{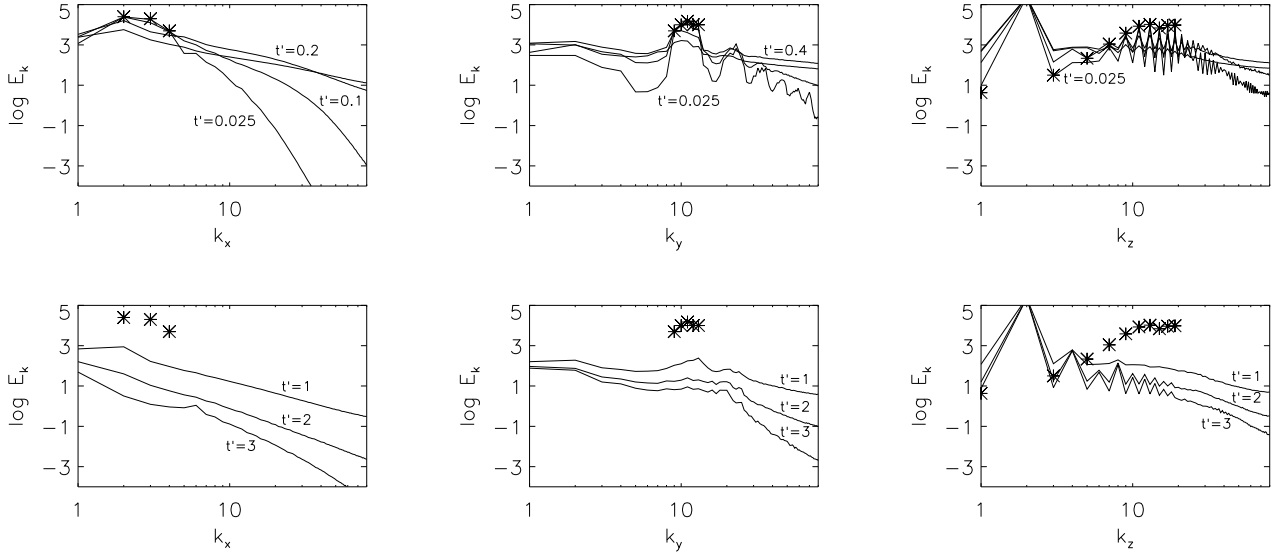}
\caption{The evolution of the field with a large perturbation. Spectra
are shown at the same times, and in the same format as in Fig.
\ref{small}.}
\label{large}
\end{figure*}

Figure \ref{large} shows the results for the large perturbations, which contain 10\%
of the background field energy in each of the top ten eigenmodes,  that is, we have forced these
eigenmodes to retain their distinct identities for amplitudes considerably
larger than they would have according to Fig. \ref{small}.  Even this
though does not allow them to remain independent in their subsequent evolution.
In contrast, the nonlinear spreading to other modes, and development of
the Hall cascade, simply proceeds even faster than in Fig. \ref{small},
until by $t'=3$ one once again obtains the standard Hall cascade, the
details of the initial conditions having been almost completely erased, 
and there certainly being no trace of any growing peaks.

\section{Discussion}

We propose that the entire concept of Hall instabilities has two
significant weaknesses even in the purely linear regime considered by R\&G.
First, it depends on having a rather special background field, satisfying Eq.
(\ref{ZeroHall}).  Second, the resulting instabilities are not
small-scale at all; they are only slightly smaller in scale than the
background itself.

Furthermore, we have demonstrated that even if one carefully
constructs the initial conditions to reproduce unstable eigenmodes of the 
background field, which may lead to small-scale instabilities, the hypothesis 
of R\&G's work, the most one can accomplish is to slightly delay the onset of
the usual Hall cascade.  None of the fully nonlinear calculations, of ourselves
or numerous other authors
\citep{biskamp96,biskamp99,dastgeer00,dastgeer03,cho04,shaikh05,cho09},
which have a broad variety of different initial conditions including
sufficiently strong magnetic fields, have ever found anything other than a 
standard Hall cascade.

\cite{cumming04} also consider the possibility of a Hall instability, and
its relevance for the evolution of neutron star magnetic fields. They
clarify the nature of the instability: a background shear in the electron
velocity drives growth of long-wavelength, i.e., {\it large-scale},
perturbations. They explicitly remark that short-wavelength modes
are unaffected. They also note that if the picture of a turbulent Hall
cascade is correct, then this Hall instability probably does not change the
long-term evolution of the field, since intermediate scales will `fill in'
as the cascade develops.  This filling-in is precisely what we have
demonstrated here.

We conclude therefore that Hall instabilities may exist in the sense
that one can do the linear stability analysis and obtain growing modes, but
if one considers the full nonlinear evolution according to Eq. (\ref{eq:A}),
one finds that these modes are completely subsumed into the standard Hall
cascade, and `instabilities' play no significant role in the dynamics of
Eq. (\ref{eq:A}).

\begin{acknowledgements}
This work was supported by the Science \& Technology Facilities Council 
[grant number PP/E001092/1]. We would also like to acknowledge the 
comments of the referee, Professor Brandenburg, who's remarks have allowed 
us to improve and clarify our manuscript.
\end{acknowledgements}

\bibliographystyle{aa}
\bibliography{Wareing}

\end{document}